# In situ visualization of tip undercooling and lamellar microstructure evolution of sea ice with manipulated orientation

*Tongxin Zhang, Zhijun Wang**, *Lilin Wang**, *Junjie Li, Jincheng Wang*

*State Key Laboratory of Solidification Processing, Northwestern Polytechnical University, Xi'an 710072, China*

**Abstract:** Sea ice growth with lamellar microstructure containing brine channels has been extensively investigated. However, the quantitative growth information of sea ice remains lack due to the uncontrolled crystalline orientation in previous investigations. For the first time, we in-situ observed the unidirectional growth of lamellar sea ice with well-manipulated ice crystal orientation and visualized tip undercooling of sea ice. A semi-empirical model was proposed to quantitatively address the variation of tip undercooling with growth velocity and salinity and compared with a very recent analytical model. With the real-time observation, interesting phenomena of doublon tip in cellular ice growth and growth direction shift of ice dendritic tip were discovered for the first time, which are attributed to the complex solutal diffusion and anisotropic interface kinetics in sea ice growth. The quantitative experiment provides a clear micro scenario of sea ice growth, and will promote relevant investigations of sea ice in terms of the theoretical approach to describing the diffusion field around faceted ice dendritic tip.

Keyword: sea ice, tip undercooling, lamellar microstructure

Dendritic growth of hexagonal ice in water remains a puzzling phenomenon and relevant studies have been particularly extensive because of its importance in many domains [1].For example, around 70% of the earth's surface is covered by ocean. In different seasons, the formation and reduction of sea ice is of great concern due to its significance in its potential ecological and geopolitical impacts [2]. At its maximum, sea ice covers 7% of the earth's surface area [3]. Unlike pure ice, frozen sea water is composed of a complicated lamellar microstructure with a network of brine channels and pores varying in size from a few micrometers to millimeters. The microstructures in sea ice under various thermal conditions and salinities can greatly affect the optical and mechanical properties of sea ice [4], which may influence our lives in various aspects such as the incident ultraviolet irradiance on snow and sea ice-covered ecological habitat [3, 5, 6], the remote-sensing of sea ice [7] and the possibility to collapse which results in easier sea-level rise [8]. Therefore, the morphology of ice crystal is of great significance for the macroscopic properties of sea ice. However, the faceted nature of ice [9] endows ice dendritic growth unique properties that distinct from non-faceted dendrites [10]. And many popular dendritic growth theories were based on non-faceted crystal, which may become invalid for faceted dendrites [11]. For example, the MSC theory [12] and the LMK theory [13] remain the most acceptable version for ice dendritic growth and have been frequently used to describe growth of basal plane ice in pure water only. For ice dendritic growth in aqueous solutions, however, Bulter [14] pointed out that solvability theory is valid only for basal tip radius. In addition, symmetry breaking of growth morphology of ice crystal usually occurs in many previous studies under both free growth [15-17] and directional growth [14] conditions, which has made ice a peculiar subject in dendritic growth. Therefore, understanding the formation of the microstructure in sea ice has been one of the general concerns in geophysics and condensed matter physics.

* Corresponding author. zhjwang@nwpu.edu.cn

* Corresponding author. wlilin@nwpu.edu.cn

The details on sea ice microstructure evolution characterized by lamellar platelets remain elusive even though water is one of the most common substances on earth. Most of the physical understanding of pattern formation in ice growth was accumulated from free growth condition [13, 16, 18-21]. Extensive investigations [13, 16, 18-22] tried to reveal the selection of ice growth morphology and tip velocity of ice dendrites for given bath undercooling. The morphologies evolution of ice can be addressed by the coupling of thermal and/or chemical diffusion and interface kinetics. However, for the confined growth of ice under imposed thermal gradient, the physical understanding of pattern formation is quite limited despite its direct relation with the formation of sea ice microstructure. Preliminary studies [23-25] have shown the elongated knife-edged interface morphology and its variations against solute concentration and growth velocity. However, the absence of orientation manipulation of the lamellar ice restricts quantitative exploration of sea ice growth because the results greatly depend on the angle between the direction of thermal gradient and crystalline orientation of ice in confined growth condition.

In confined growth of sea ice, the most important parameters for theoretical consideration are tip undercooling, tip radius and primary spacing for given thermal gradient and pulling velocity. For example, the tip undercooling, primary spacing and tip radius are usually measured with given thermal gradient and pulling velocity in directional solidification to reveal the theory of solidification via non-facet transparent materials like succinonitrile (SCN) alloys [26, 27]. Up to now, few of these parameters has been quantitatively presented in previous investigations on sea ice growth due to some great challenges. The first challenge is the preparation of a single ice crystal with well-controlled orientation. The quantitative investigation of interface microstructure evolution needs a single ice crystal with its basal plane parallel to the direction of pulling velocity and perpendicular to the observation plane. The second challenge is the precise measurement of temperature at the ice dendritic tip. The tip undercooling measurement requires a thermal resolution higher than 0.01K, and it is impossible to quantitatively measure the temperature with a thermocouple in the micro scale.

Here we successfully manipulate the ice orientation in a thin channel and precisely in-situ measure the tip undercooling of an advancing solid/liquid (S/L) interface. The variation of tip undercooling with pulling velocity and initial salinity is quantitatively revealed. With these quantitative experimental data, a semi-empirical model for tip undercooling selection is proposed and compared to up-to-date analytical model [28]. The methods and results are of great help for understanding the nature of microstructure evolution of sea ice.

The experiments were performed in a unidirectional freezing manner with each single ice crystal of the same crystal orientation grown in a capillary tube. **Figure 1** shows the control principle of crystal orientation with birefringence of ice and the schematic diagram of unidirectional freezing with measurement of tip undercooling. It was reported by Brewster [29, 30] that ice is a “positive uniaxial doubly-refracting” material due to its hexagonal crystal symmetry (see **Fig. 1(a)**), and any incident light whose direction is not parallel to the C-axis of ice crystal can be “resolved into ordinary and extraordinary components” through ice crystal [31]. It has been shown in *Physics of ice* [31] that “the birefringence of ice is extremely useful for determining the orientations of the C-axis in crystals and observing the grain structure in thin sections of poly-crystalline ice”.

Based on crystal optics, the specific orientation relation of an ice crystal with respect to the laboratory frame ***A-P-L*** (***A-A*** is the direction of analyzer, ***P-P*** is the direction of polarizer and ***L*** is the direction of incident polarized light) is directly linked with the dimensionless intensity $I_{\perp}/I_0$ of incident polarized light, which is determined by both α and β with three relations Eq.1-3, as illustrated in **Fig. 1(b)**.

$$\text{for } 0 \le \alpha \le 90° : \Delta N_{\alpha} = \sqrt{\frac{1}{\frac{\cos^2 \alpha}{N_0{}^2} + \frac{\sin^2 \alpha}{N_e{}^2}}} - N_0 \qquad \text{(Eq. 1)}$$

$$R = \Delta N_{\alpha} \cdot d \qquad \text{(Eq. 2)}$$

$$I_{\perp}/I_0 = \sin^2(\frac{\pi R}{\lambda}) \cdot \sin^2 2\beta \qquad \text{(Eq. 3)}$$

where α "tilt angle of optical axis" is an acute angle of C-axis of ice crystal tilting from the incident light direction ***L***; β "extinction angle" is an angle between the projection line of C-axis in ***A-P*** plane and the direction ***P-P***; $N_0$ and $N_e$ are refractive indexes for ordinary and extraordinary light through ice; $\Delta N_{\alpha}$ is the birefractive index of ice with a corresponding $\alpha$; $R$ is the optical path difference of ordinary and extraordinary light due to the birefringence of ice; $d$ is the thickness of ice crystal being transmitted; $I_0$ is the incident intensity of polarized light and $I_{\perp}$ is the transmitted intensity of polarized light; $\lambda$ is the wavelength of polarized light.

The $\Delta N_{\alpha}$-α curve from Eq.1 is plotted in **Fig. 1(c),** where $\Delta N_{\alpha}$ monotonically increases with α (0≤α≤90°) to a maximum value with a corresponding position labeled as "*M*". According to Eq.3, the dimensionless intensity $I_{\perp}/I_0$ which corresponds to the length of line "*OF*" against extinction angle β on polar coordinate system exhibits a quartic symmetry, as plotted in **Fig. 1(d)**. When β = 0°, 90°, 180° or 270°, $I_{\perp}/I_0$ = 0, extinction will occur in which ice sample appears dark and such direction is called "extinction direction". α and β can be manipulated by changing the position of the specimen box fixed to the ***X-Y-Z*** frame where the ice crystal grows with respect to the ***A-P-L*** frame. By manipulating the two parameters α and β of the ice crystal to specific values based on the orientation relation between the frame

***X-Y-Z*** and the laboratory frame ***A-P-L***, the dark "extinction position" (noted as "*E1*") can be determined and one can finally obtain an ice crystal whose C-axis is perpendicular to both the thermal gradient and incident light for the following directional freezing experiments. A step-by-step methodology based on crystal optics is graphically illustrated in **Fig. 1(e-j)** [32]. A single ice crystal was guaranteed by uniformly dark image in every step when rotating the specimen under crossed polarizers because grain boundaries of poly-crystalline ice can be recognized if the specimen does not appear uniformly dark in the extinction position [31].

**Figure 1(k)** shows the schematic diagram of horizontal directional freezing stage and measurement of tip undercooling $\Delta T_{tip}$ by differential visualization (DV) method [33, 34]. Rectangular glass capillary tube with an inner space dimension of 0.05 × 1 mm$^2$ (VitroCom brand) was adopted as sample cell for unidirectional freezing of PVA solutions. In each capillary tube prior to in-situ directional freezing, the C-axis of single ice crystal was manipulated to be perpendicular to directions of both the thermal gradient and incident light. The imposed temperature gradient for directional growth was in the range of $G$ = 5.00 ± 0.80 K/mm. The microstructure evolution of S/L interface was recorded by a CCD camera. In addition, the ice crystal orientation was simultaneously detected through a pair of polarizers to guarantee that the crystal orientation remained unchanged during freezing of NaCl solutions. Tip undercoolings under different pulling velocities for all samples were precisely obtained by DV method [33, 34].

In the experiments, the single ice crystal whose orientation was specially manipulated as the edge plane is unidirectional solidified within a capillary tube in NaCl solutions with comparable salt concentrations to sea water under an imposed temperature gradient. The growth velocity $V$ and initial solute concentrations $C_0$ are the key variables to be controlled for the variation of tip undercooling. In-situ observations of S/L interface microstructure evolution of the edge plane ice with various morphologies are obtained. By using DV method, tip undercoolings of partially- and fully-developed lamellar substructure under various growth conditions can be precisely measured.

The unidirectional solidification of single ice crystals into modeled sea water, i.e. NaCl solutions composed of ultrapure water (provided by deionizer when the resistance of water comes to 18.25 MΩ) and NaCl solute (AR, 99.5 %) degassed under vacuum condition. Five solute concentration (0.05 M, 0.1 M, 0.2 M, 0.3 M, 0.6 M) and four step-increment pulling velocity regimes (V1, V2, V3 and V4, from slow to fast) are performed by means of the capillary technique under a well-defined and fixed temperature gradient (G = 5.00 ± 0.80 K/mm). The same procedure was performed for other samples with different initial solute concentration $C_0$ under four pulling velocity regimes that was altered in a step-increment manner.

**Supplementary Movie 1** demonstrated DV method of tip undercooling measurement in real space. Variations of tip undercooling with pulling velocity $V$ and initial solute concentration $C_0$ were summarized in **Fig. 2(a)**. It shows that the tip undercooling increased with the increase of the salinity. Without pulling, the S/L interface keeps the undercooling of $-m_L C_0$ ($m_L$ is the liquidus slope of NaCl aqueous solution and $C_0$ is the initial solute concentration of NaCl solute. The $m_L$ was precisely measured as shown in the **Appendix B**). Once the interface moves, the solutal boundary layer builds up ahead of the S/L interface, causing constitutional undercooling. As the pulling velocity increases, the tip undercooling gradually decreases and approaches to $-m_L C_0$. The precise tip undercooling can be analyzed to further reveal the growth mechanism of sea ice growth.

The tip undercooling in directional solidification is composed of three parts,

$$\Delta T_{tip} = \Delta T_c + \Delta T_r + \Delta T_k \approx \Delta T_c \quad \text{(Eq. 4)}$$

where $\Delta T_c$, $\Delta T_r$ and $\Delta T_k$ are the contribution of solutal, curvature and kinetic effects, respectively. The kinetic term $\Delta T_k$ is usually very small for edge plane of ice [35], and by considering the kinetic coefficient [36], the kinetic undercooling does not exceed 0.01 K (the growth velocity is in the order of $10^{-5} m/s$ and the kinetic attachment coefficient $K_T$ is in the order of $10^{-3} m/(s \cdot K)$). The curvature effect is difficult to be considered due to the faceted nature of ice tip [12, 17, 37]. As reported previously, the tip radius is crystal orientation dependent——the tip radius is different for basal plane and edge plane in free growth conditions [12], and some researchers [13, 21] tried to evaluate the tip radius of ice dendrites by two distinct tip radii ($R_1$ and $R_2$) for basal and edge plane. In this study, we found that the "apparent tip radius" be scattered in a narrow range of 5.10-9.05 μm and the curvature undercooling do not exceeds 0.03K for all solidification samples (see **Appendix A**), which is minor compared with the measured tip undercooling. Hence the measured tip undercooling is mainly constitutional by the build-up of solute boundary layer in front of the S/L interface. Then the rule that governs constitutional undercooling will be a key to understand the microstructure evolution in sea ice growth.

Although ice is of faceted nature, the solute pile-up at the ice dendritic tip still obeys the partition rule and the diffusion equation in the liquid. In the following, a model is established to account for the ice tip undercooling results. To better elucidate the tip undercooling model, a schematic diagram is given in **Fig. 3**. The coordinate $X$ is a distance coordinate in the direction of pulling velocity $V$. The variable $C$ in relation to $X$ gives the concentration profile of the liquid near the ice dendrite, which consists of the concentration profile of an inter-dendritic liquid region (in dark blue solid line) and that of a liquid region beyond the dendritic tip (in red solid line). The liquid concentration $C_L(x)$ at any given distance $x$ beyond the ice dendritic tip can be split into two parts ($\overline{C}_L(x)$ and $\int C_L(x)$) [38] combined with Zener's assumption [39] as shown in **Fig. 3**. $\overline{C}_L(x)$ and $\int C_L(x)$ are the solute diffusion parallel to the growth direction and lateral solute diffusion related to the ice dendritic tip morphology, respectively. Usually, the lateral diffusion is described via analogy of sphere growth with tip radius as the characteristic diffusion length in non-faceted system [38]. However, the tip radius of the faceted ice is almost unchanged as the pulling velocity increases within our experimental range, and tip radius is no longer validate as a characteristic length in addressing the effect of lateral diffusion. Different from traditional analysis, an effective diffusion length $L_{eff}(V)$ is proposed here to replace the tip radius in the lateral diffusion undercooling term. The $L_{eff}(V)$ is to be determined. Derivation is started with the steady-state diffusion equation in a moving frame (directional solidification at speed $V$) in the following

$$D_L \nabla^2 C_L + V \cdot \nabla C_L = 0 \qquad \text{(Eq. 5)}$$

where $D_L$ is the diffusion constant for solute, $C_L$ is the solute concentration as a function of its position. The solute balance at the interface gives

$$V_n(1-k_0)C_I + \left(\frac{\partial C}{\partial n}\right)\Bigg|_{C=C_I} = 0 \qquad \text{(Eq. 6)}$$

where $V_n$ is the velocity normal to the S/L interface, $C_I$ is the solute concentration at the S/L interface, $k_0$ is the equilibrium distribution coefficient for ice crystal, $\left(\frac{\partial C}{\partial n}\right)\Big|_{C=C_I}$ is the solute concentration gradient normal to the S/L interface.

In one dimensional coordinate $X$ ( $X$ is the distance coordinate in the direction of growth velocity $V$, see **Fig. 3**) with an additional far field solute concentration of $C_0$, Eq. 5 and Eq. 6 become

In the liquid phase: $D_L \frac{\partial^2 C_L}{\partial x^2} + V \cdot \frac{\partial C_L}{\partial x} = 0$ (Eq. 7)

Solute mass balance at the S/L interface: $\frac{\partial C_I}{\partial x} = -\frac{V}{D_L}(1-k_0)C_I$ (Eq. 8)

Far field solute concentration: $C(x=\infty) = C_0$ (Eq. 9)

Similar to the investigation of Burden and Hunt [38], by defining $x$ as the distance beyond the ice dendritic tip, then the solute concentration in the vicinity of the ice dendritic tip with a given $x$ can be assumed as (see **Fig. 3**)

$$C_L(x) = \overline{C}_L(x) + \int C_L(x) \qquad \text{(Eq. 10)}$$

Here $\overline{C}_L = \overline{C}_L(x)$ is the solute buildup in the vicinity of ice dendritic tip for a planar S/L interface; $\int C_L = \int C_L(x)$ is the radial solute build-up in the vicinity of ice dendritic tip and is also related to the latter defined effective diffusion length $L_{eff}$. Here it should be noted that the symbol '$\int$ ' on the left hand side of '$\int C_L(x)$' does not represent any integral.

The $\overline{C}_L(x)$ decays exponentially with varying $x$ and must satisfies

$$D_L \frac{\partial^2 \overline{C}_L}{\partial x^2} + V \cdot \frac{\partial \overline{C}_L}{\partial x} = 0 \quad \text{(Eq. 11)}$$

Giving the relation

$$\frac{\partial \overline{C}_L}{\partial x} = \frac{V}{D_L}(C_0 - \overline{C}_L) \quad \text{(Eq. 12)}$$

And at the tip where $X = X_t(x=0)$, $\frac{\partial \overline{C}_L}{\partial x}$ equals the concentration gradient in the inter-dendritic liquid region behind the tip as suggested by Burden and Hunt [38] (see **Fig. 3**)

$$\frac{\partial \overline{C}_t}{\partial x} = \frac{V}{D_L}(C_0 - \overline{C}_t) = \frac{G_L}{m_L} \quad \text{(Eq. 13)}$$

where $G_L$ is the temperature gradient for directional solidification of ice and $m_L$ is the liquidus slope of NaCl solutions. For a planar S/L interface, $\overline{C}_L$ is sufficient to take away rejected solute from the S/L interface. However, when breakdown of the planar S/L interface into ice dendrites occurs, $\int C_L(x)$ is additionally needed to take away solute from the ice dendritic tip (see **Fig. 3**).

At the tip where $X = X_t(x=0)$, Eq. 10 becomes

$$C_t = \overline{C}_t + \int C_t \quad \text{(Eq. 14)}$$

where the subscript "$t$" refers to the ice dendritic tip, differentiating Eq. 14 gives

$$\frac{\partial C_t}{\partial x} = \frac{\partial \overline{C}_t}{\partial x} + \frac{\partial \int C_t}{\partial x} \quad \text{(Eq. 15)}$$

The first term on the RHS of (Eq. 15) is the magnitude of concentration gradient at the tip [40], which obeys an exponential decay with increasing $x$ beyond the ice dendritic tip. The second term on the RHS of Eq. 15 is the magnitude of concentration gradient via radial diffusion around the faceted ice tip with an effective diffusion length $L_{eff}$ and was previously obtained by using Zener's "approximate" method [39],

$$\frac{\partial \int C_t}{\partial x} = -\frac{\int C_t}{L_{eff}} \quad \text{(Eq. 16)}$$

Combining (Eq. 8 and Eq. 13-16), we have

$$\frac{G_L}{m_L} - \frac{\int C_t}{L_{eff}} = -\frac{V}{D_L}(1-k_0)(\overline{C}_t + \int C_t) \quad \text{(Eq. 17)}$$

Rearranging terms

$$\int C_t \left[ -\frac{1}{L_{eff}} + \frac{V}{D_L}(1-k_0) \right] = -\frac{G_L}{m_L} - \frac{V}{D_L}(1-k_0)\overline{C}_t \quad \text{(Eq. 18)}$$

By substituting $\frac{G_L}{m_L}$ in Eq. 18 using Eq. 13 and rearranging terms, we have

$$\int C_t = \frac{\frac{V}{D_L}(k_0-1)C_0 - k_0 G_L}{-\frac{1}{L_{eff}} + \frac{V}{D_L}(1-k_0)} \quad \text{(Eq. 19)}$$

It should be noted that in Eq. 18 we no longer use the approximation $\frac{V}{D_L}(1-k_0) \ll \frac{1}{L_{eff}}$ which might lead to wrong estimation of tip concentration. The main task is to find the rule that governs constitutional undercooling $\Delta T_c$. Substituting Eq. 13 and Eq. 19 into the constitutional undercooling $\Delta T_c$ in Eq. 4 and considering the extremely low solubility of NaCl [41-43] in ice crystal lattice ($k_0 \approx 0$) gives

$$\Delta T_c = m_L(C_0 - C_t) = m_L(C_0 - \overline{C}_t) - m_L \int C_t = \frac{D_L G_L}{V} + m_L C_0 \frac{\frac{V}{D_L}}{-\frac{1}{L_{eff}} + \frac{V}{D_L}} \quad \text{(Eq. 20)}$$

By introducing the characteristic diffusion length $L_c = \frac{D_L}{V}$, Eq. 20 becomes

$$\Delta T_c = L_c \cdot G_L - m_L C_0 \frac{1}{\frac{L_c}{L_{eff}} - 1} \quad \text{(Eq. 21)}$$

The first term on the RHS of Eq. 21 is related to the solute diffusion parallel to the growth direction which is equivalent to a planar S/L interface. The second term on the RHS of Eq. 21 corresponds to the constitutional undercooling by the radial diffusion around the ice dendritic tip and is denoted as $\delta\Delta T = -m_L C_0 \frac{1}{\frac{L_c}{L_{eff}} - 1}$.

Therefore, two diffusion terms ($\overline{C}_t$ and $\int C_t$) can approximately describe the tip undercooling of ice dendrites as shown in **Fig. 3**. Thus the total tip undercooling of the ice dendrite is

$$\Delta T_{tip} \approx \Delta T_c = L_c(V) \cdot G_L + \delta\Delta T = L_c(V) \cdot G_L - m_L C_0 \cdot \frac{1}{\frac{L_c(V)}{L_{eff}(V)} - 1} \qquad \text{(Eq. 22)}$$

Here $D_L$ was calculated as $D_L = 2/(1/D_{Na^+} + 1/D_{Cl^-})$ [44] ($D_L$ was taken as a value of $7.737 \text{x} 10^{-10} m^2/s$ near 0°C [45]). And $-m_L C_0$ is the freezing point depression of NaCl solutions with an initial solute concentration of $C_0$ and a liquidus slope $m_L$ precisely measured by cooling-curve method (see **Appendix B**). The second term $\delta\Delta T = -m_L C_0 \cdot \frac{1}{\frac{L_c(V)}{L_{eff}(V)} - 1}$ corresponds to a more precise form of the constitutional undercooling caused by radial diffusion around the ice dendritic tip and is denoted as $\delta\Delta T$. The effective diffusion length $L_{eff}(V)$ can be calculated based on the precise results in **Fig. 3(a)** via Eq. 22. **Fig. 3(b)** gives the variation of $L_{eff}(V)$ against the characteristic diffusion length $L_C(V)$. It is interesting to notice that $L_{eff}(V) \approx 0.5 L_c(V)$ for all solute concentrations. Based on the experiment results, $L_{eff}(V)$ is growth velocity-dependent and in the same order of magnitude of $L_c$. The magnitude of the effective diffusion length $L_{eff}(V)$ is comparable to that of the characteristic diffusion length $L_c(V)$ while in non-faceted dendritic growth, the characteristic lateral diffusion length is much smaller than $L_{eff}(V)$ [38, 46, 47].

With $L_{eff}(V) \approx 0.5 L_c(V)$, $\Delta T_c$ will approach $-m_L C_0$, then the total tip undercooling $\Delta T_{tip}$ will approach $-m_L C_0$ at large pulling velocity. The proposed model can represent the change of tip undercooling at different initial salinity and pulling velocity. The experimental results and the proposed model showed that the freezing point depression at the ice dendritic tip decreased to $-m_L C_0$ in freezing of sea water at large pulling velocity, indicating that there were no obvious solute build-up at the tip. This was consistent with the prediction of the effective distribution coefficient (i.e. $k_{eff}$) widely used in sea water solidification [48]. When the growth velocity is large enough [49], $k_{eff}$ will approach unity and the S/L interface will experience "partitionless" solidification along the growth direction. The lateral diffusion will vanish, and the lamellar microstructure remains well-developed.

Very recently, Alexandrov and Galenko [28] has provided analytical solutions for two- and three-dimensional growth of angled dendrites and arbitrary parabolic/paraboloidal dendrites by solving the corresponding boundary integral equations. Their model is also helpful in predicting the variation of measured tip concentration against different ice dendritic tip morphology with various initial solute concentration and growth velocities in this work. The boundary integrals $I_C^{\xi}$ of solute diffusion for two-dimensional angled dendrites and arbitrary parabolic/paraboloidal dendrites were considered in this work. At the vertex of an angled dendrite ($|x|$ is small enough) with an interface function of $\xi(x) = a|x| + b$, the dimensionless tip concentration $C_t / C_0$ was proved to satisfy

$$\frac{C_t}{C_0} = \frac{1}{1 - \frac{2}{\pi}\arctan\left|\frac{1}{a}\right|} \quad \text{(Eq. 23)}$$

which is termed as angular tip model in the following. And for an arbitrary parabolic dendrite with an interface function of $\xi(x, y) = ax^2 + bx + c (a < 0)$, $C_t / C_0$ was proved to satisfy a relation with varying $a$

$$\frac{C_t}{C_0} = \frac{1}{1 - \sqrt{\frac{\pi P_c}{2a}} \exp(-\frac{P_c}{2a}) erfc(\sqrt{-\frac{P_c}{2a}})} (a < 0) \quad \text{(Eq. 24)}$$

which is termed as 2D Ivantsov model in the following, where $P_c$ is the solutal Péclet number.

The dimensionless tip concentrations $C_t / C_0$ of all samples were plotted against $a$ and compared to the predictions of Eq. 23 and Eq. 24 as shown in **Fig. 4**. The variable "$a$" is related to the measured tip morphology in our work. Three types of ice dendritic tip morphology were observed and named as "cellular", "symmetry breaking cellular" (SBcellular in short) and "angular", respectively. It was interesting that, for the data points within the range of $a \geq 1.5$, angular tip model (Eq. 23) can well predict the results regardless of the difference in tip morphology. When $a \leq 1.5$, the data points became more and more scattered around the prediction of angular tip model. On the contrary , 2D Ivantsov model (Eq. 24) was invalid in predicting the ice dendritic tip concentration within the whole range of $a$ in this work, which was most probably due to the faceted nature of ice dendrites that made its shape always not precisely parabolic. To be more specific, when growth velocity increased, $a$ increased to a threshold value of around $a \approx 1.5$, and the ice dendritic tip experienced a transition from cellular tip to angular tip, which corresponded to solute diffusion-controlled growth to interface kinetics-controlled growth. Similar morphology transition from parabolic tip to angular tip was also reported by Brener & Temkin [50] and Galenko et al. [51]. In their discussions, anisotropic interface kinetics were claimed to be the main factor of this morphology transition of a growing dendritic tip. And their discussions can be helpful in addressing the variation of ice dendritic tip morphology in our work.

Our semi-empirical model can well describe the variation of ice dendritic tip undercooling by introducing the effective diffusion length $L_{eff}$. And the analytical models by Alexandrov and Galenko [28] were also proved to be valid for data points that satisfied $a \geq 1.5$. Owing to the fact that both the ice dendritic tip morphology and tip undercooling varied with growth conditions, we further speculated that there should be a geometry dependence of $L_{eff}$ on $a$. Hence, we coupled the main conclusions of our model (Eq. 20 and Eq. 21) and the angular tip model (Eq. 23) to reveal the physical relation between $a$ and $L_{eff}$. It can be proved that substitution of the tip concentration $C_t$ into Eq. 20 and Eq. 21 with Eq. 23 with newly introduced variable $\Delta T_f = m_L C_0$ and $f(a) = \arctan\left|\frac{1}{a}\right|$ will yield

$$L_{eff}(a) = (1 - \frac{1}{\frac{G \cdot L_c}{\Delta T_f} + \frac{1}{f(a)-1} + 2}) \cdot L_c \qquad \text{(Eq. 25)}$$

Therefore, from Eq. 25 it can be seen that, for faceted ice dendrite, the effective diffusion length is also a function with respect to its tip morphology.

The quantitative measurement of the tip undercooling has provided important information about tip undercooling selection and the build-up of solute boundary layer in front of the ice dendritic tip with given thermal gradient and pulling velocity. Besides, the morphology evolutions in the well-designed experiments also show intriguing phenomena for understanding the microstructure evolution of sea ice. **Fig. 5** shows the S/L interface morphologies of 0.1 M (a-d) and 0.6 M (e-h) NaCl solutions at different pulling velocities. The morphologies were chosen at the steady-state solidification. The speed-up videos (see **Supplementary Movie 2** and **3**) showed the morphology evolution with the increase of pulling velocity for 0.1 M and 0.6 M samples, respectively.

For the 0.1M system, there was a dynamic adjustment of the primary lamellar spacing at the lowest pulling velocity as shown in **Fig. 5 (a-c)**. Although the tip position was almost fixed, the primary spacing was far from steady state. After a long interval of observation, the morphology was confirmed to be oscillatory unstable with unevenly spaced lamellar arrays of ice dendrites. The new tip was generated via tip splitting instead of overgrowth of side-branches. The tip splitting behavior of the edge plane is interesting, which conflicts the tip stability of the edge plane of ice in free growth conditions as reported by Koo et al [12]. There was a solute boundary layer ahead the interface in directional growth at very low pulling velocity as shown in the measurement of tip undercooling. Then the tip splitting is controlled by solute diffusion. The doublon cellular tip in **Fig. 5 (a)** is similar to other well-investigated systems like PEO-SCN [52] (cubic, rough), Xenon [53] (cubic, rough) and biphenyl [54] (monoclinic, weakly faceted). Here the in-situ observation indicates an instable growth mode of sea ice with low salinity and enriches the experimental insight with doublon tip behavior of faceted ice. As the pulling velocity further increased, uniformly spaced lamellar arrays of ice dendrites were observed as shown in **Fig. 5 (d)**. In fact, similar to the morphology of ice dendrites **Fig 5 (a)**, cellular ice tips with knife-edged shape were also frequently reported in free growth of ice [12]. The specific physical origin of knife-edged ice dendritic tip remains unclear. In the non-facet dendritic growth, the tip is usually symmetric. On the contrary, for ice dendritic growth, the dendritic tip usually appears asymmetric [12, 16, 17, 21, 55], which is also termed as “symmetry breaking of an ice dendrite”.

**Figure 5 (d-f)** showed that the morphologies of the 0.6M system were dendritic. The increased salinity enlarged the constitutional undercooling and induced the well-aligned dendritic arrays. All the dendritic tips appeared triangular in the vicinity of the dendritic vertex. In the dendritic growth theories, the dendritic tip is greatly affected by the crystalline orientation with anisotropy. Generally, the dendritic tip will grow along with the preferred orientation in the anisotropic crystalline [56-59]. In this study, however, as the pulling velocity increased, the dendrite tip showed an obvious shift of growth direction as shown in **Fig. 5 (e-h)** even when the direction of heat flux is parallel to the preferred orientation. In this study, the crystalline orientation with basal plane parallel to the directions of both the thermal gradient and incident light remained unchanged, which was confirmed by polarized light. The shift of the growth direction of the ice dendritic tip with increased pulling velocity was very strange and has never been reported in directional growth of sea ice. The mechanism may be related to the growth kinetics of different crystallographic planes and need to be further revealed. In directional growth of sea ice, we can not draw a clear conclusion about the continuous change of ice dendrites growth direction since in *The Chemical Physics of Ice* [1], relevant remarks can be found as "The growth mechanism is not entirely clear but may involve some sort of stepped or segmented growth whose pattern depends upon the different temperature variation of growth velocities parallel to C- and A- axes". Here we have only preliminary assumption based on previous studies—in free growth from undercooled water, there are plenty of experiment results and relevant explanations [60-62]. As previously reported by Macklin & Ryan [60, 61] and Lindenmeyer &

Chalmers [62], for ice grown freely with different bath undercoolings in both pure water and aqueous solutions, the ice growth morphology will experience a transition from coplanar growth (rounded disk) to non-coplanar growth ("symmetrically situated pyramidal segments") at a certain critical undercooling. A "step growth mechanism" was proposed, which claimed that the actual growth direction of ice dendritic tip can be decomposed into two growth velocities parallel and perpendicular to the basal plane. Because the growth kinetics of edge and basal plane are different due to the different nature of the two interfaces, different bath undercoolings will yield different growth rates [63] for edge and basal plane, and the actual growth direction of ice dendritic tip will deviate from basal plane to some extent, depending on the undercooling and solute additives. Therefore, in our study, it is reasonable to speculated that the change of ice dendritic tip growth direction can also be qualitatively addressed by the "step growth mechanism". In addition, the solute impurity modified the deviation effect of ice dendritic tip growth direction, which was also qualitatively consistent with the results of Pruppacher [64] and Macklin & Ryan [63].

In conclusion, complex microstructure evolution of sea ice with well controlled orientation was in-situ observed in a unidirectional manner. Precise measurement of tip undercooling has been made for edge plane S/L interface at different initial salinities under various growth velocities. In an experimental consideration, it is revealed for the first time that the effective diffusion length near a faceted ice tip is comparable to that of characteristic diffusion length instead of the tip radius, which is different from non-faceted systems. With a proposed semi-empirical model, the solutal profile at the ice tip can be well described. And measured sea ice tip concentrations were in good agreement with the very recent analytical angular tip model when $a \geq 1.5$. Besides, the physical relation between the effective diffusion length and the ice dendritic tip morphology was further revealed. Moreover, for the first time, the tip splitting behavior of the edge plane and the shift of ice tip growth direction have been directly observed in directionally solidified sea ice, which reveals the complex interactions between the solutal diffusion-controlled and interface kinetics-controlled growth for sea ice growth.

## Acknowledgements

This work was supported by the National Key R&D Program of China (Grant No.2018YFB1106003), National Natural Science Foundation of China (Grant No. 51701155), and the Fundamental Research Funds for the Central Universities (3102019ZD0402).

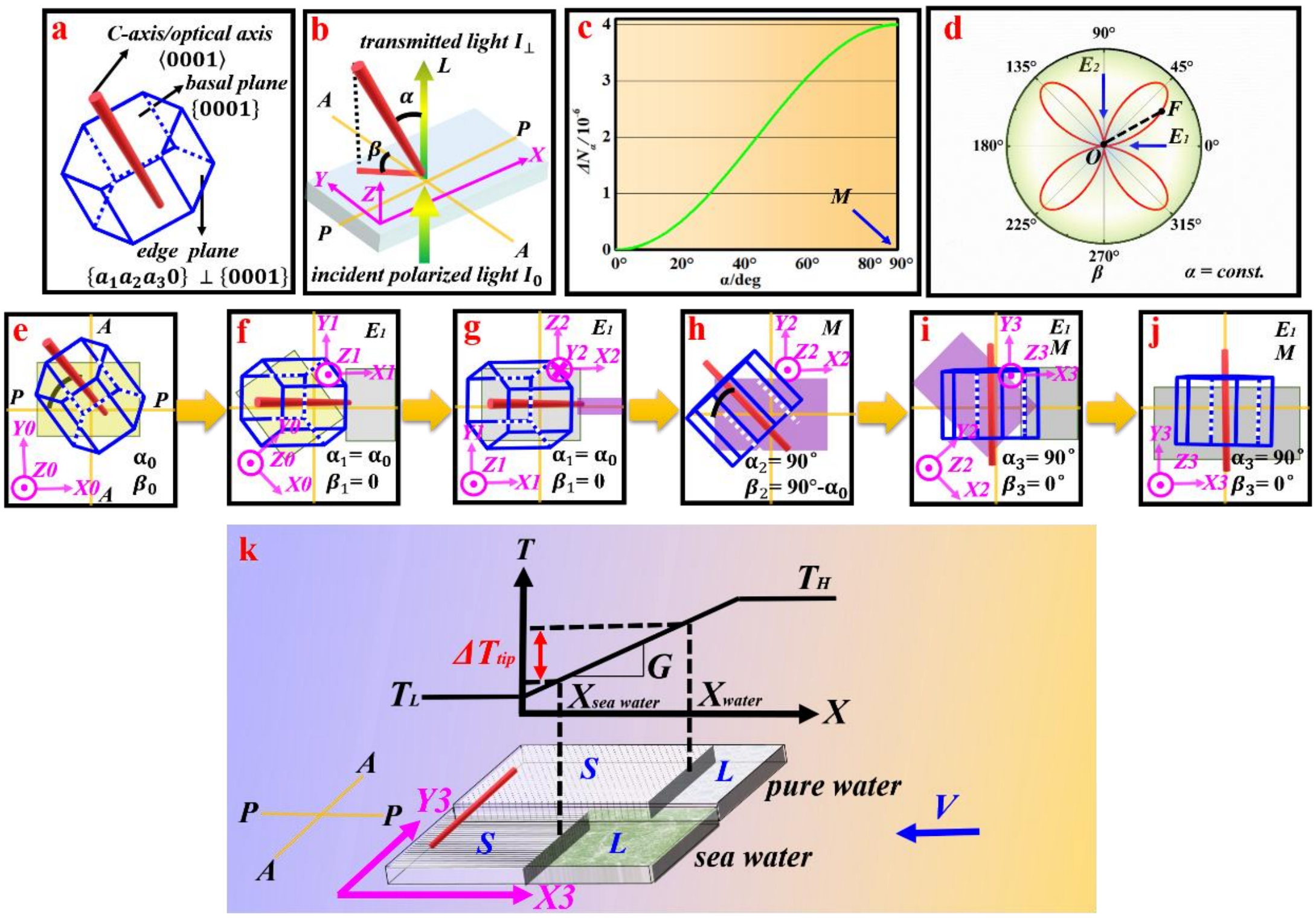

**FIG. 1** The schematic description of orientation manipulation of ice crystal and S/L interface undercooling measurement. **FIG** (**a-d**) represent the physical foundation of the orientation detection based on crystal optics. **FIG** (**e-j**) show the specific operations for manipulating a single ice crystal with designed orientation in a series of rectangular glass tubes. **FIG** (**k**) is the principle for tip undercooling measurement. In details: (**a**) Description of the crystal orientation of a single ice crystal via the relation between its basal plane $\{0001\}$, edge plane $\{a_1a_2a_30\}$ and its optical or C axis that is represented by a red rod parallel to <0001> direction; (**b**) The orientation relation between the frame ***X-Y-Z*** tied to a glass tube where a single crystal ice is grown and the laboratory frame ***A-P-L*** (***A-A*** is the direction of analyzer, ***P-P*** is the direction of polarizer and ***L*** is the direction of incident polarized light with an incident intensity of $I_0$ and a different transmitted intensity of $I_\perp$ due to the birefringence of ice). Two parameters α and β are defined as follows: α ("tilt angle of optical axis") is defined as the angle between the optical axis and the incident light with direction ***L*** in the laboratory frame ***A-P-L***. β ("extinction angle") is defined as the angle between the projection line of optical axis and ***P-P*** in plane ***A-P***. α and β can be manipulated by changing the position of the frame ***X-Y-Z*** with respect to the frame ***A-P-L***; (**c**) The $\Delta N_\alpha$-α curve showing that $\Delta N_\alpha$ monotonically increases with α (0⩽α⩽90°) to a maximum value according to Eq.1 with a corresponding position labeled as "*M*"; (**d**) Variation of the dimensionless intensity $I_\perp/I_0$ with a corresponding length of line "*OF*" against extinction angle β on polar coordinate system exhibiting a quartic symmetry according to (Eq.3): when β = 0°, 90°,

180° or 270°, $I_{\perp}/I_0 = 0$, extinction will occur in which ice sample appears dark and such direction is called "extinction direction". For convenience in the following demonstration, the corresponding positions where β = 0° and 90° are labeled as "*E1*" and "*E2*", respectively; (**e**) The orientation relation between the frame $\boldsymbol{X_0}$-$\boldsymbol{Y_0}$-$\boldsymbol{Z_0}$ that is tied to the first specimen where a single ice crystal with arbitrary orientation is grown and the frame ***A-P-L*** is $\alpha_0$, $\beta_0$; (**f**) The orientation relation between the first (on the left) and the second (on the right) specimen after rotating the first specimen to position "*E1*", making the projection line of optical axis of ice in plane ***A-P*** parallel to direction ***P-P***, so that the orientation relation between the frame $\boldsymbol{X_1}$-$\boldsymbol{Y_1}$-$\boldsymbol{Z_1}$ that is tied to the second specimen and the frame ***A-P-L*** is $\alpha_1 = \alpha_0$, $\beta_1 = 0°$ (position "*E1*"); (**g**) The orientation relation between the second (on the left) and the third (on the right) specimen, with plane $\boldsymbol{X_2}$-$\boldsymbol{Y_2}$ of the frame $\boldsymbol{X_2}$-$\boldsymbol{Y_2}$-$\boldsymbol{Z_2}$ placed perpendicular to plane $\boldsymbol{X_1}$-$\boldsymbol{Y_1}$ of the frame $\boldsymbol{X_1}$-$\boldsymbol{Y_1}$-$\boldsymbol{Z_1}$, making the optical axis of ice lie in plane $\boldsymbol{X_2}$-$\boldsymbol{Y_2}$ in the third specimen; (**h**) The optical axis of ice is made to lie in plane ***A-P*** by rotating the third specimen in (**g**) about direction ***P-P*** by 90° so that the orientation relation between the frame $\boldsymbol{X_2}$-$\boldsymbol{Y_2}$-$\boldsymbol{Z_2}$ that is tied to the third specimen and the frame ***A-P-L*** is $\alpha_2 = 90°$ (position "*M*"), $\beta_1 = 90° - \alpha_0$; (**i**) The orientation relation between the third (on the left) and the fourth (on the right) specimen, with the optical axis of ice parallel to direction ***A-A*** by rotating the frame $\boldsymbol{X_2}$-$\boldsymbol{Y_2}$-$\boldsymbol{Z_2}$ that is tied to third specimen to position "*E1*" so that the orientation relation between the frame $\boldsymbol{X_3}$-$\boldsymbol{Y_3}$-$\boldsymbol{Z_3}$ that is tied to the fourth specimen and the frame ***A-P-L*** is $\alpha_3 = 90°$ (position "*M*"), $\beta_3 = 0°$ (position "*E1*"); (**j**) The orientation of each ice crystal grown in capillary tubes in this paper simultaneously satisfies positions "*M*" and "*E1*" as shown in the fourth specimen; (**k**)Schematic diagram of horizontal directional freezing stage and measurement of tip undercooling $\Delta T_{tip}$ by differential visualization method. The crystal orientations of ice in two glass tubes for directional growth are the same as represented by the red solid rod lying within plane ***X-Y*** and parallel to the ***Y*** axis in which the laboratory frame ***A-P-L*** coincides with the frame $\boldsymbol{X_3}$-$\boldsymbol{Y_3}$-$\boldsymbol{Z_3}$.

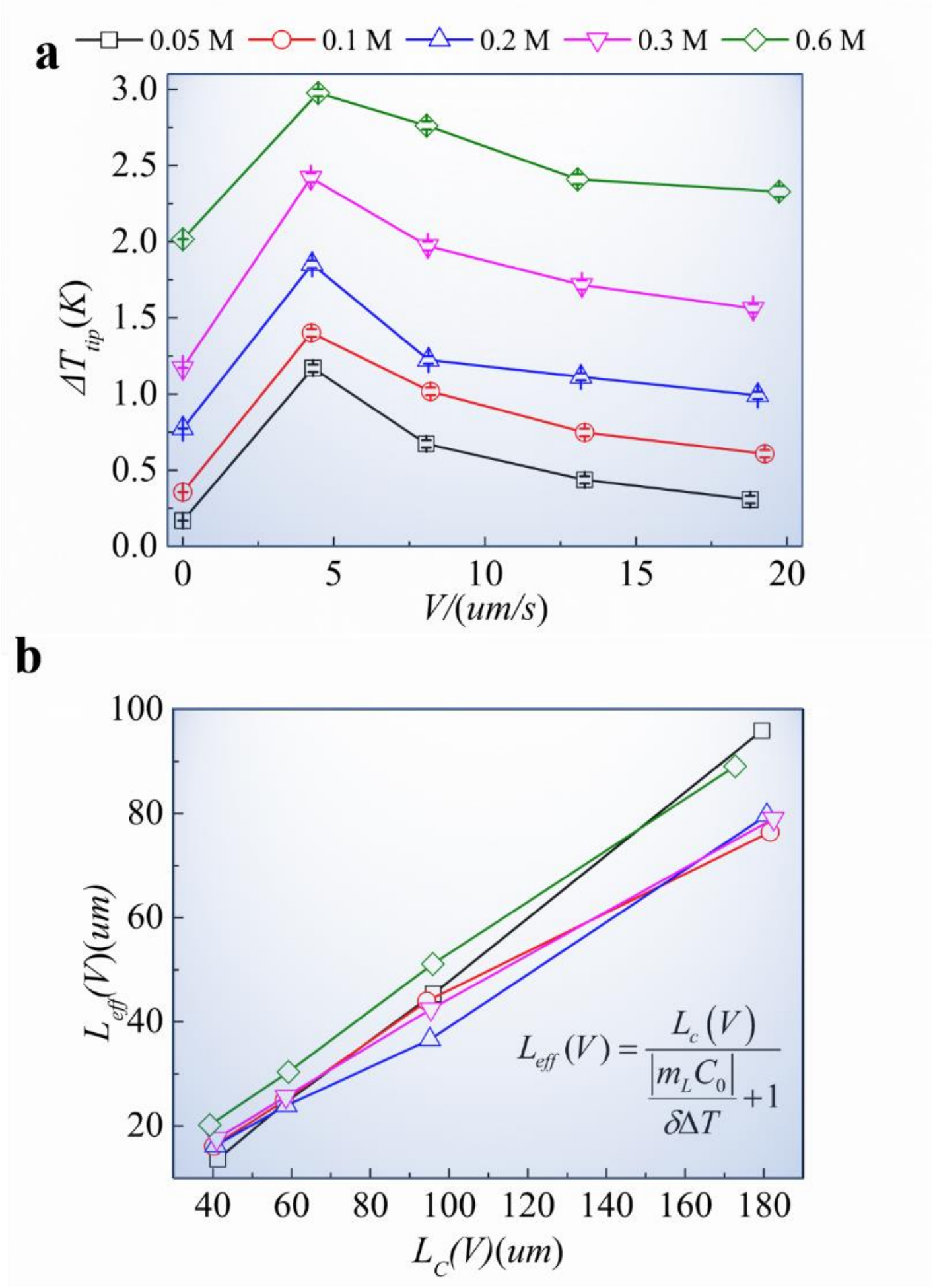

**FIG. 2 (a)** Variation of tip undercooling $\Delta T_{\mathrm{tip}}$ under steady-state solidification of ice for different pulling velocities and different initial solute concentrations $C_0$. The error bar for each undercooling data is 0.024 K. $\Delta T_{\mathrm{tip}}$ at 0 μm/s corresponds to static S/L interface undercooling (i.e. $-m_L C_0$) for all concentrations precisely measured by cooling curve method (see **Appendix B**). **(b)** The effective diffusion length $L_{eff}(V)$ calculated according to our semi-empirical model *vs* the characteristic diffusion length $L_C(V)$ for different initial solute concentrations $C_0$ based on the result of **FIG. 2(a)**.

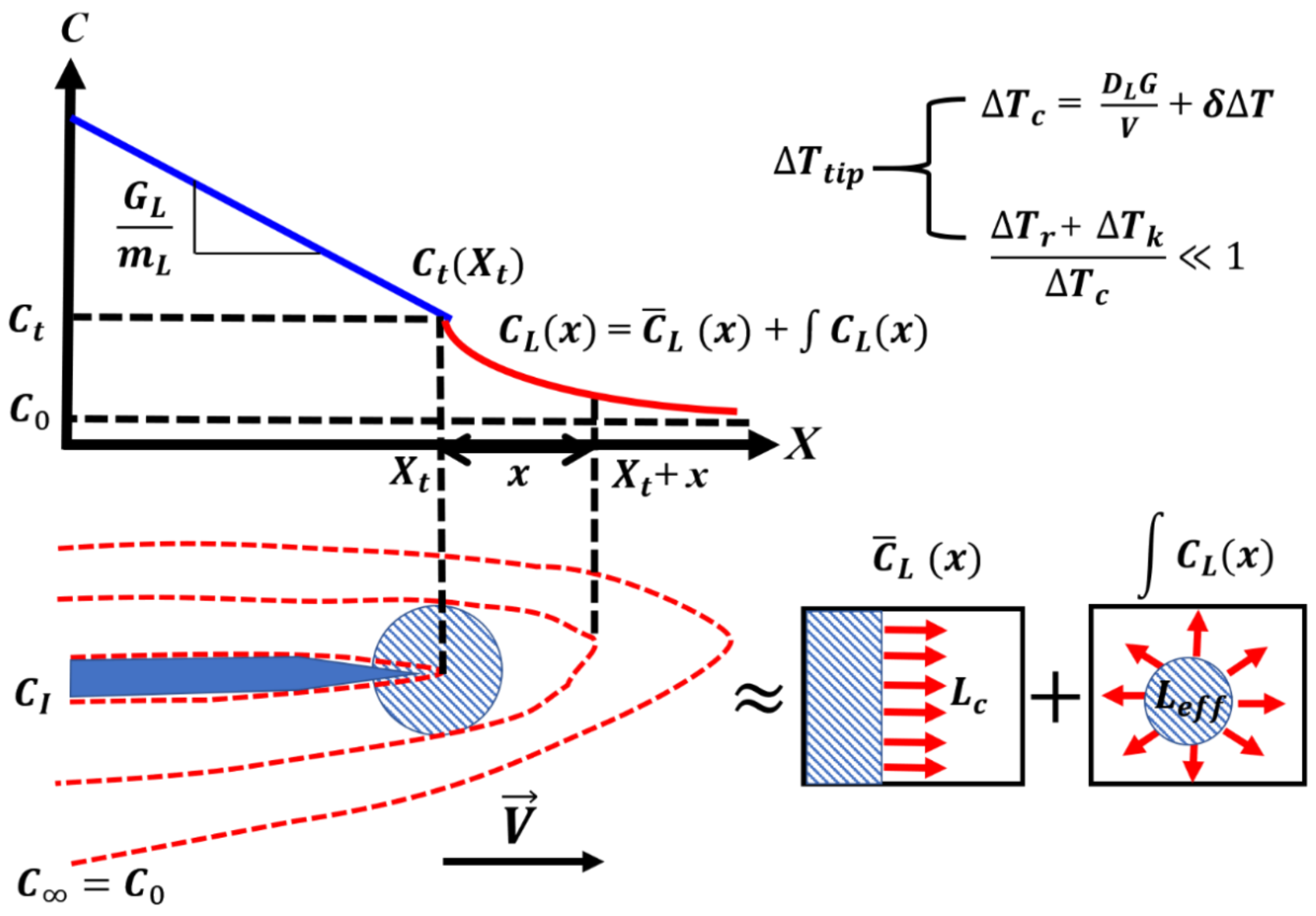

**FIG. 3** The graphical illustration of our semi-empirical model in this paper. The tip undercooling of ice dendrite in this paper is proved to be mainly constitutional. The faceted ice dendritic tip is drawn with solute concentration contours in red dash lines (the innermost line in close proximity to the faceted ice dendrite represents the concentration contour of the S/L interface $C_I$ that equals $C_t$ at $X = X_t(x = 0)$ ; the outermost one represents the concentration contour that equals far field concentration $C(x = \infty) = C_0$ ). A shaded circle representing the radial diffusion field with a radius of effective diffusion length $L_{eff}$ is drawn around an ice dendritic tip.

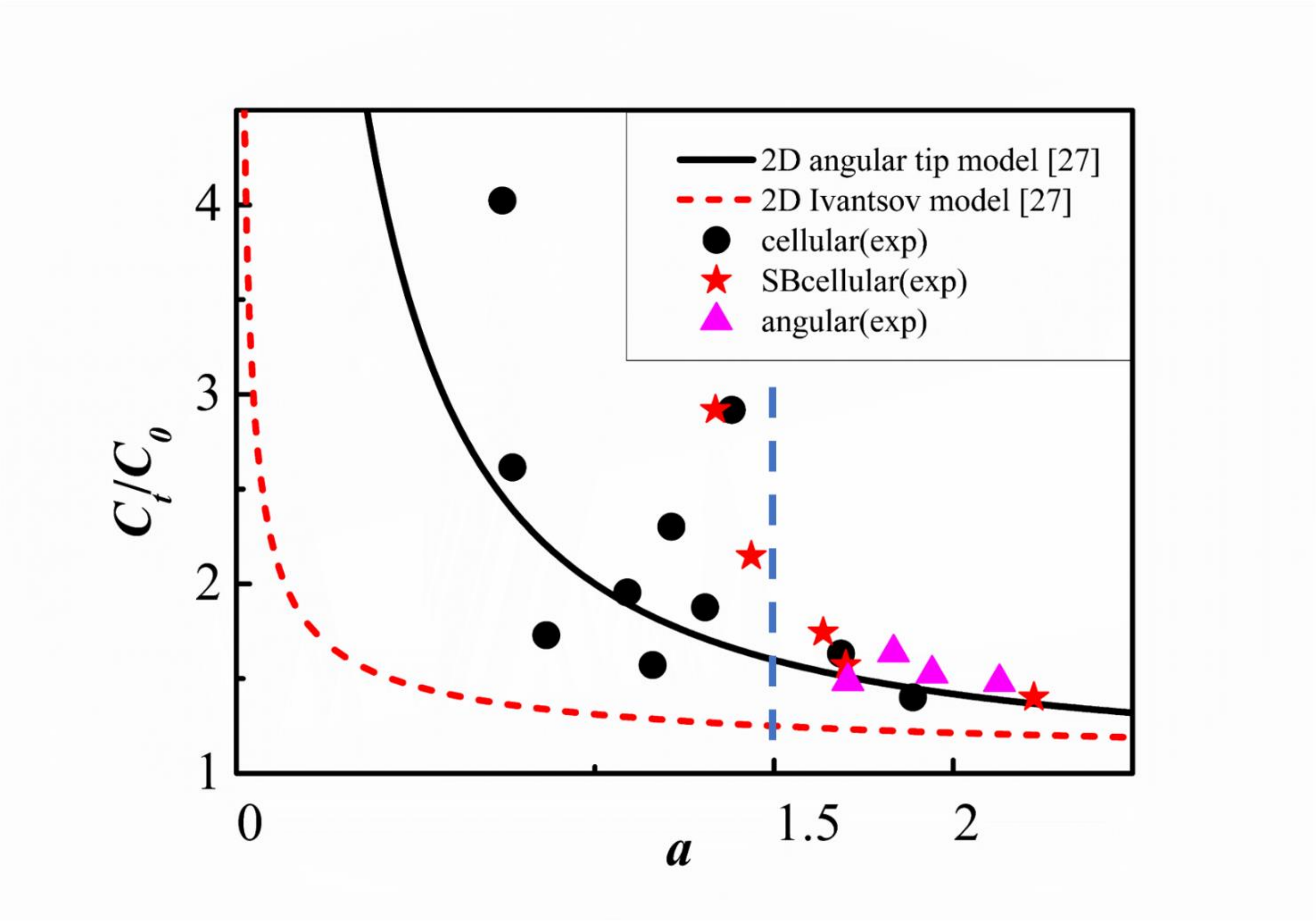

**FIG. 4** Variation of measured dimensionless ice tip concentration $C_t/C_0$ against $a$ that is related to the measured tip morphology in our work. The measured dimensionless ice tip concentration $C_t/C_0$ were obtained by converting tip undercooling results in **Fig.2 (a)** into $C_t/C_0$ via Eq. 4 and Eq. 20. Data points were plotted with different labels according to different types of ice tip morphology, which include "cellular" (black solid Circle), "SBcellular" (red solid Star) and "angular" (magenta solid UpTriangle), respectively. Angular tip model (red dash line) and 2D Ivantsov model (black solid line) with varying $a$ were simultaneously plotted.

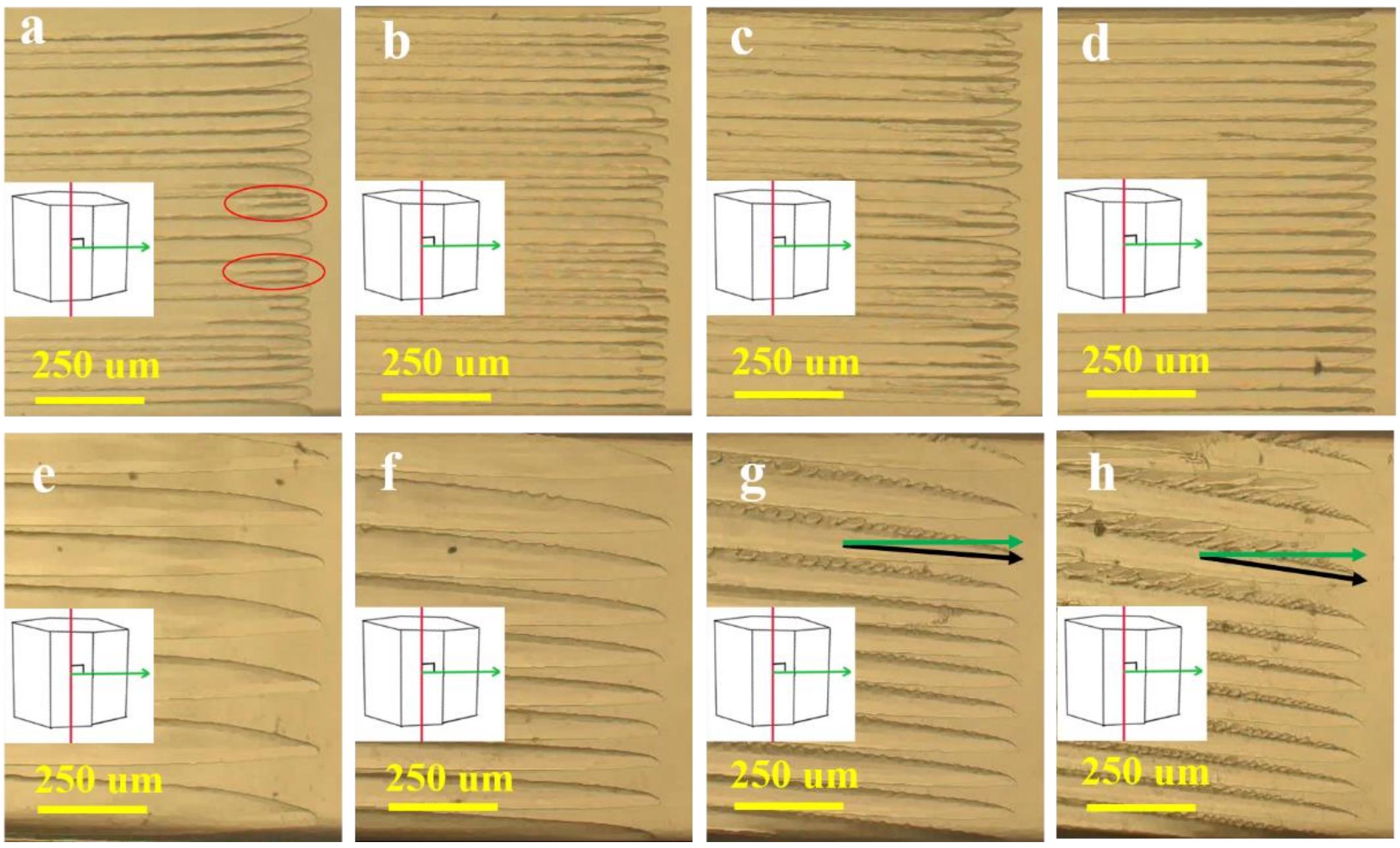

**FIG. 5 (a)-(d)** The S/L interface morphology evolution with increasing pulling velocity for $C_0$ = 0.1 M **(a)** Doublon cellular tip morphology (marked in red circle) was observed during steady-state solidification, which showed the coexistence of asymmetrical cellular and knife-edged lamellar microstructures. The pulling velocities $V$ were 4.26 μm/s, 8.20 μm/s, 13.30 μm/s and 19.26 μm/s for **(a)**, **(b)**, **(c)** and **(d)**, respectively. $G$ = 5.39 K/mm for **(a)**, **(b)**, **(c)** and **(d)**. **(e)-(h)** The S/L interface morphology evolution with increasing pulling velocity for $C_0$ = 0.6 M. **(g)-(h)** Deviation of tip actual growth direction (arrow in black) from pulling velocity direction (arrow in green) of fully-developed lamellar morphology with distinct side-branches on one side from the direction of basal plane (see the arrow in green in the inset, which parallel to the growth direction). The pulling velocities $V$ are 4.48 μm/s, 8.07 μm/s, 13.07 μm/s and 19.74 μm/s for **(e)**, **(f)**, **(g)** and **(h)**, respectively. $G$ = 4.80 K/mm for **(e)**, **(f)**, **(g)** and **(h)**. The inset shows the orientation relation of ice crystal with respect to the growth cell, in which the red arrow represents the c-axis of ice and the green arrow represents the direction of pulling velocity.

## Supplementary materials

**Supplementary Movies: The typical movies during the in-situ investigations.**

**Appendix A: The estimation of curvature undercooling in this work**

The following part addresses the estimation of the curvature undercooling $\Delta T_r$ in (A10) based on our experimental results. The 0.6 M sample was chosen for the following estimation because it is easy to produce the most finest tips at this solute concentration according to our experiment results and correspond to the maximum possible curvature undercooling. By selecting several points (squared dots in magenta) on the edge of the tip for a parabolic fit ( $y = C + B_1 x + B_2 x^2$ )(**Fig. A1**), the tip radius is estimated for all pulling velocities (**Fig. A2**) and summarized in **Table A**. All results of tip radius are found to be scattered within the narrow range of 5.10 - 9.05 μm/s. It should be noted that although it is difficult to obtain the realistic tip morphology in 3D case, the "apparent tip radius" mentioned here via a parabolic fit is sufficient for the estimation of $\Delta T_r$ in terms of its order of magnitude. Besides, from other preliminary studies [21, 55, 65, 66], $\Delta T_r$ is speculated to be either smaller or in the same order of magnitude compared to our estimation since their smallest measured tip radius is either larger [12, 55] or comparable [21] to ours, which indicates the same order of magnitude of $\Delta T_r$.

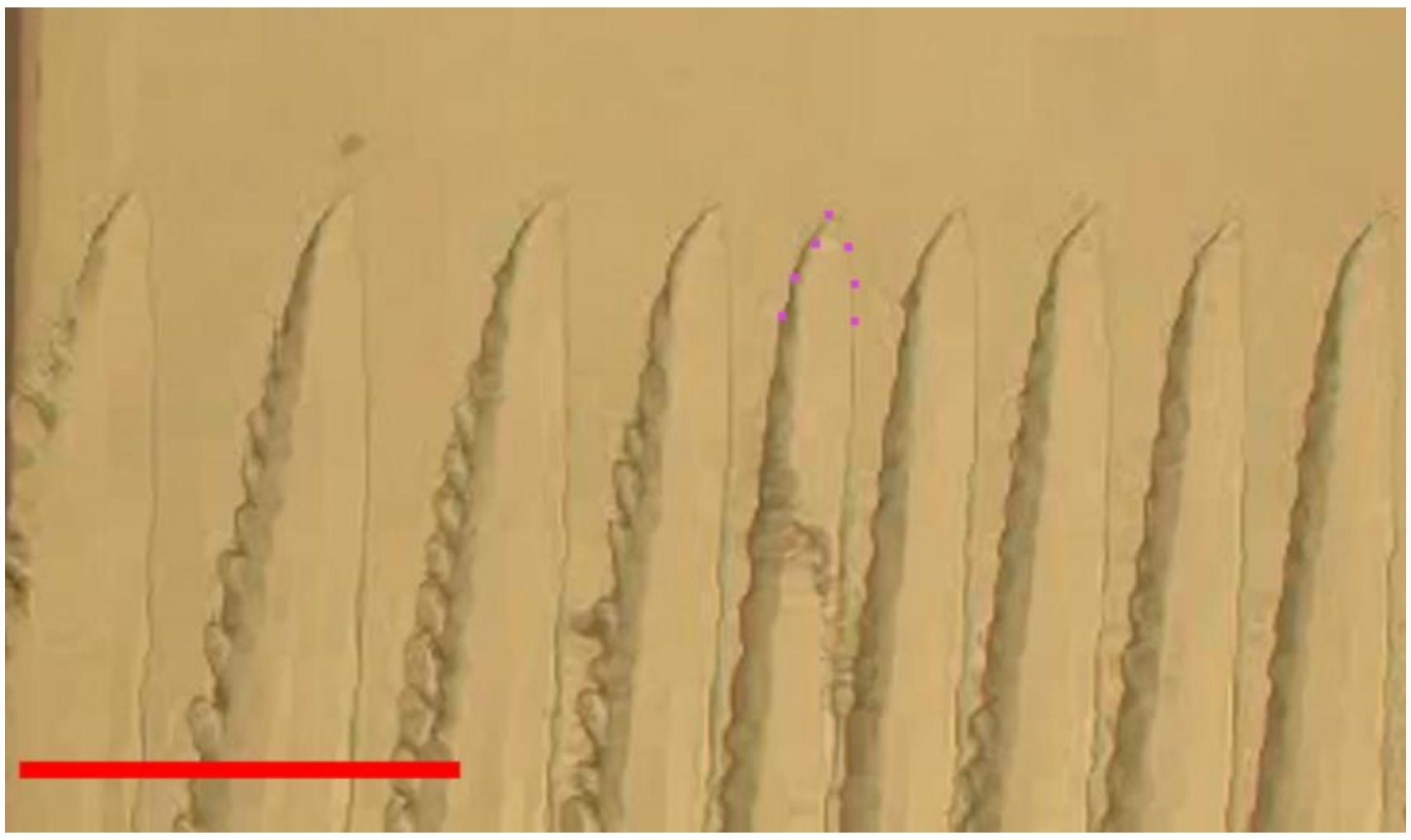

**Fig. A1** Tip morphology of 0.6 M sample whose $V$ = 13.07 μm/s and $G$ = 4.796 K/mm. By selecting several points (squared dots in magenta) on the edge of the tip for a parabolic fit, the tip radius is estimated at the tip. Each square in dash line has an edge length of 108.5 μm in this figure. The image is enhanced to exhibit clearer tip morphology. The scale bar is 250 μm in this figure.

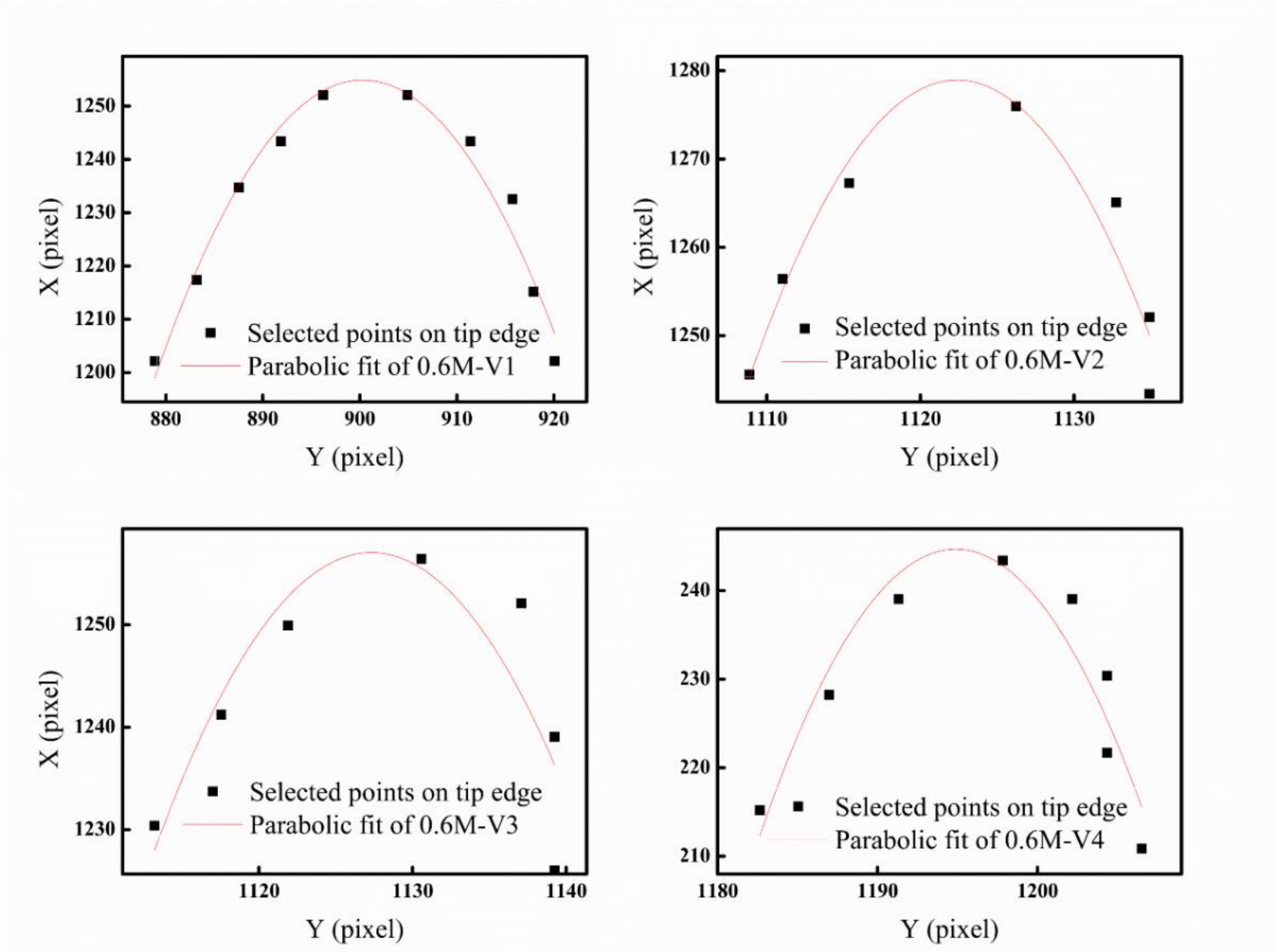

**Fig. A2** Parabolic fit of tip morphology for all pulling velocities (V1 = 4.48 μm/s, V2 = 8.07 μm/s, V3 = 13.07 μm/s and V4 = 19.74 μm/s) of 0.6 M sample in this study.

**Table A**

| V | B2 | error | tip radius/μm | error | curvature/μm$^{-1}$ | error |
|---|---|---|---|---|---|---|
| V1 | 0.122 | 0.00854 | 9.05 | 0.001 | 0.111 | 0.0171 |
| V2 | 0.184 | 0.03177 | 5.97 | 0.002 | 0.168 | 0.0635 |
| V3 | 0.146 | 0.04478 | 7.55 | 0.002 | 0.132 | 0.0896 |
| V4 | 0.216 | 0.03597 | 5.10 | 0.003 | 0.196 | 0.0719 |

For convenience at this point, we take the lowest value of the tip radius as $R = 2\,\mu$m μm ($\kappa = 0.5\,\mu$m$^{-1}$) which is lower than any of the value in **Table A**.

$$\Delta T_r = \frac{\gamma_{SL} T_L}{L} \cdot \kappa \qquad \text{(A1)}$$

where $\gamma_{SL}$, $T_L$, $L$ and $\kappa$ are ice-water interfacial energy, melting point of ice, latent heat of ice and curvature of the ice dendrites tip, respectively. By taking the value of $\gamma_{SL}$ = $51\times10^{-3} J/m^3$ [67], $T_L = 273.15K$, $L$ = $3.06\times10^{8} J/m^3$ and $\kappa$ = $0.5\times10^{6} m^{-1}$ in SI, $\Delta T_r$ **= 0.023 K**, which is the same order of magnitude as the experimental error of tip undercooling measurement in this work. Therefore, it can be concluded that the curvature undercooling $\Delta T_r$ is minor compared to the constitutional component $\Delta T_c$ in our study. Hence the compositional undercooling is major in terms of the total tip undercooling.

## Appendix B: The freezing point depression of NaCl sample used in this work

Results of measured freezing point depression of NaCl solution in the concentration range of up to 6.7 wt.% (equivalent to 1.2 M) with high precision of 0.001 $K$ by cooling curve method are shown in **Fig. B**. Our results are in close agreement with previous reports [68, 69]. The liquidus slope $m_L$ was measured to be 0.592 $K$/wt.%.

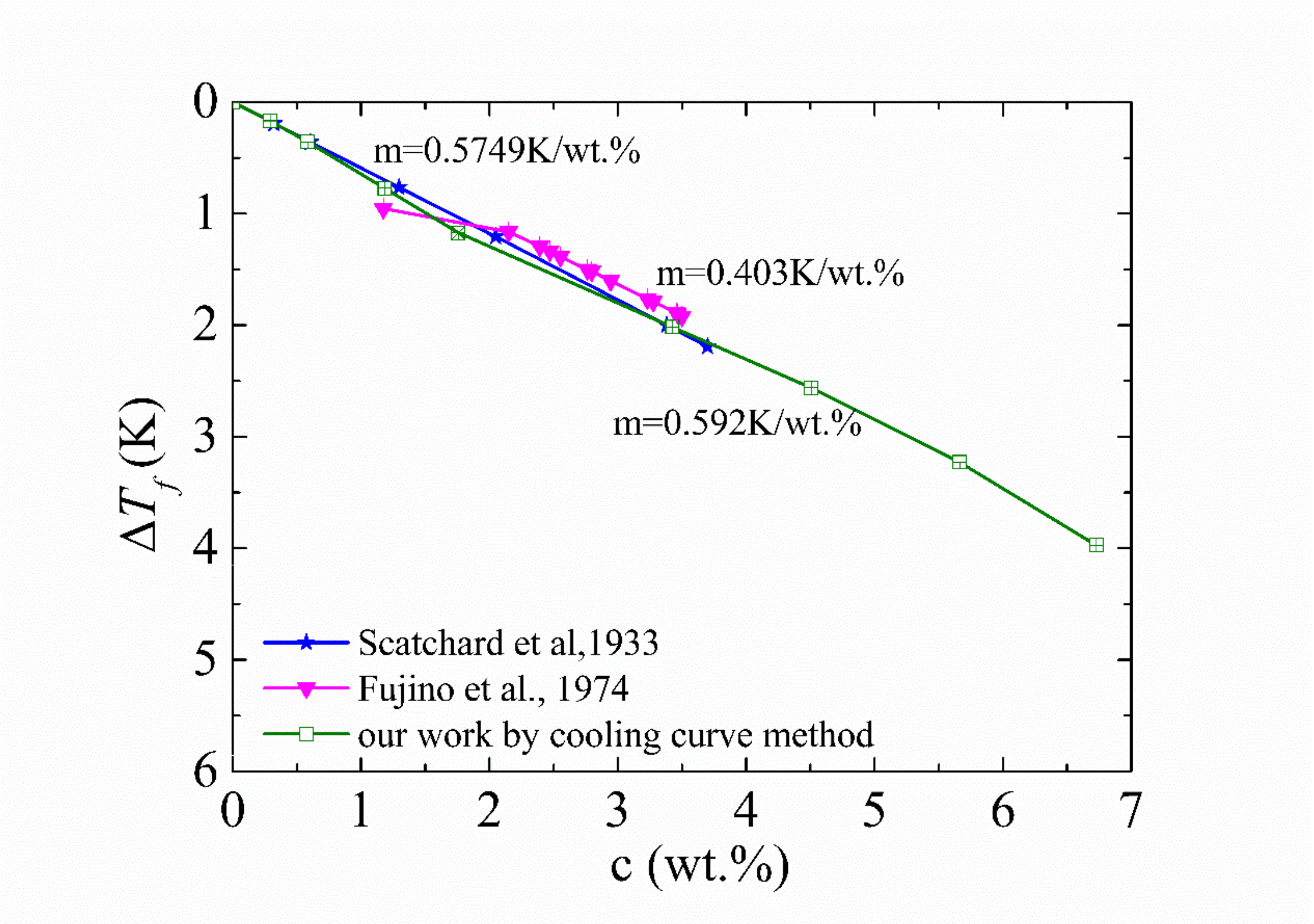

**FIG. B** Freezing point depression of our NaCl sample within the compositional range of up to 6.7 wt.% (equivalent to 1.2 M) by cooling curve method. The results are compared with previous reports by Fujino et al.[68] and Scatchard et al.[69].

## Supplementary movies

**Movie 1**: A demonstration movie of "differential visualization method" used in our study;

**Movie 2**: A speed-up movie which includes doublon tip morphology at low pulling velocity (*V* = 4.26 μm/s) and further S/L interface morphology evolution at higher pulling velocities (*V* = 8.20 μm/s, 13.30 μm/s and 19.26 μm/s) for 0.1 M sample (*G* = 5.39 K/mm);

**Movie 3**: A speed-up movie of S/L interface morphology evolution for 0.6 M sample which demonstrates the gradual deviation of tip growth direction from basal plane direction with the increment of pulling velocity (*V* = 4.48 μm/s, 8.07 μm/s, 13.07 μm/s and 19.74 μm/s ; *G* = 4.80 K/mm).